\documentclass[doublecol]{epl2}

\title{Resolving social dilemmas on evolving random networks}
\shorttitle{Resolving social dilemmas on evolving random networks}

\author{Attila Szolnoki\inst{1} \and Matja{\v z} Perc\inst{2}}
\shortauthor{Attila Szolnoki and Matja{\v z} Perc}

\institute{\inst{1}Research Institute for Technical Physics and Materials Science, P.O. Box 49, H-1525 Budapest, Hungary\\
\inst{2}Faculty of Natural Sciences and Mathematics, University of Maribor, Koro{\v s}ka  cesta 160, SI-2000 Maribor, Slovenia}


\abstract{We show that strategy independent adaptations of random interaction networks can induce powerful mechanisms, ranging from the Red Queen to group selection, that promote cooperation in evolutionary social dilemmas. These two mechanisms emerge spontaneously as dynamical processes due to deletions and additions of links, which are performed whenever players adopt new strategies and after a certain number of game iterations, respectively. The potency of cooperation promotion, as well as the mechanism responsible for it, can thereby be tuned via a single parameter determining the frequency of link additions. We thus demonstrate that coevolving random networks may evoke an appropriate mechanism for each social dilemma, such that cooperation prevails even by highly unfavorable conditions.}

\begin{document}

\maketitle

\section{Introduction}
Social dilemmas constitute situations where the collective wellbeing is at odds with individual success. Frequently studied within the framework of evolutionary game theory \cite{msmith}, they provide some of the most challenging environments for the sustenance of cooperative behavior. As successors of evolutionary games on regular grids \cite{nowaknat92, lindgrenpd94, szabopre98}, evolutionary games on complex networks have been the subject of intense investigations in the last years \cite{szabopr07}, and their ability to promote or sustain cooperative behavior in different types of social dilemmas has often been confirmed \cite{abramsonpre01, vainsteinpre01, holmepre03, santosprl05, duranpd05, vukovpre05, santospnas06, kupermanepjb08, wangpre08}. Apart from some notable early exceptions \cite{ebelpre02, zimmermannpre04} the majority of these studies considered static interaction networks underlying the main evolutionary process. The evolution \textit{on} networks is, however, increasingly often accompanied also by the evolution \textit{of} networks \cite{holmeprl06, grossprl06, pachecoprl06, lipre07, fupre08b, szolnokiepl08, poncelaplos08, segbroeckbmc08, fupre08, segbroeckprl09}, and this not just in the context of evolutionary game theory. Indeed, networks are to be seen as evolving or adaptive entities that may substantially influence any dynamical process that is taking place on them \cite{grossjrsi08}.

Coevolutionary processes in general, \textit{i.e.} processes that happen alongside the main evolution of strategies, are currently in the focus of attention within evolutionary game theory, and it has been shown that they may very effectively promote cooperation (see \textit{e.g.} \cite{szolnokiepl08, fupre08, segbroeckprl09}). Notably, the subject of coevolution need not be the interaction network, but the term may refer to the teaching activity and related reproduction capability \cite{szolnokinjp08}, or the ability of the players to move on the spatial grid \cite{vainsteinjtb07, helbingacs08, helbingepjb09}, as well. An important observation in many cases is that coevolutionary rules may lead to highly heterogeneous states in a spontaneous manner \cite{garlaschellinat07}. Since heterogeneity has often been found favorable for the evolution of cooperation \cite{percpre08, santosnat08}, it is considered a key outcome of coevolutionary processes that lead to enhanced levels of cooperation.

Here we aim to show that simple coevolutionary rules affecting the interaction network may lead, not just to heterogeneous states promoting the cooperative strategy, but also to new dynamical processes that positively affect the evolution of cooperation. In particular, we start with a random interaction network and introduce a coevolutionary rule entailing both deletions of existing and additions of new links between players. While existing links are deleted whenever a player adopts a new strategy or its degree exceeds a threshold value, new links are added after each given number $\tau$ of game iterations. The latter parameter thus defines a time scale for the addition of new links, which may be tuned faster or slower according to the deletion of existing links. Irrespective of the time scale separation \cite{sanchezprl06, szolnokiepjb09} between them, the counteraction of deletions and additions of links largely preserves the initial random topology of the network and its heterogeneity, so that the reported ability of resolving social dilemmas is due to the spontaneous emergence of the Red Queen mechanism and group selection, which appear spontaneously in dependence on $\tau$ and the social dilemma governing the evolution of the strategies. We thus show that simple strategy independent coevolutionary rules may evoke appropriate dynamical mechanisms that affect the adoption of strategies on the macroscopic level of evolutionary games so that the governing social dilemma is resolved in favor of the collective wellbeing.

The remainder of this letter is organized as follows. First, we describe the considered social dilemmas and the protocol for the coevolution of random networks. Next we present the results, whereas lastly we summarize and discuss their implications.

\section{Social dilemmas and setup}
In what follows we consider all three major social dilemma types where players can choose either to cooperate or defect, whereby mutual cooperation yields the reward $R$, mutual defection leads to punishment $P$, and the mixed choice gives the cooperator the sucker's payoff $S$ and the defector the temptation $T$. Adopting previously introduced parametrization \cite{santospnas06}, thus designating $R = 1$ and $P=0$ as fixed, the remaining two payoffs can occupy $-1 \leq S \leq 1$ and $0 \leq T \leq 2$, where if $T>R>P>S$ we have the prisoner's dilemma game, $T>R>S>P$ yields the snowdrift and $R>T>P>S$ the stag-hunt game. Irrespective of the governing social dilemma, initially each player $x$ is designated either as a cooperator $(s_x=C)$ or defector $(s_x=D)$ with equal probability, and is placed on a random network that is constructed from $N$ individuals with an average degree $k_{avg}=4$. As usual, duplicate links are omitted. Evolution of the two strategies is performed in accordance with the Monte Carlo simulation procedure comprising the following elementary steps. First, a randomly selected player $x$ acquires its payoff $p_x$ by playing the game with all its $k_x$ neighbors. Next, one randomly chosen neighbor of $x$, denoted by $y$, also acquires its payoff $p_y$ by playing the game with all its $k_y$ neighbors. Last, if $p_x > p_y$ player $x$ tries to enforce its strategy $s_x$ on player $y$ in accordance with the probability $W(s_x \rightarrow s_y)=(p_x-p_y)/b k_q$, where $k_q$ is the largest of the two degrees $k_x$ and $k_y$, as used before in case of heterogeneous interaction topologies \cite{santosprl05, santospnas06, szolnokiepl08}. In accordance with the random sequential update, each player is selected once on average during a full Monte Carlo step.

\begin{figure}
\center \scalebox{0.5}[0.5]{\includegraphics{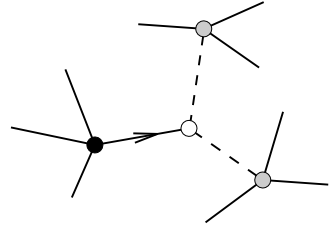}}
\caption{Schematic presentation of the coevolutionary rule affecting the interaction network. Player depicted by the black circle passes strategy to its neighbor depicted by the white circle. Consequently, the invaded player looses all its links, except from the one with the donor of the new strategy. This is marked by the dashed lines extending towards other neighboring players depicted by grey circles.}
\label{fig1}
\end{figure}

In addition to the evolution of the two strategies by each considered social dilemma, a rule for the adaptation of random interaction networks is implemented. First, whenever player $x$ adopts a new strategy all its links, except from the one with the donor of the new strategy, are deleted. This process of strategy adoption and simultaneous link deletion is demonstrated schematically in Fig.~\ref{fig1}. Hence, in addition to adopting a new strategy, the player is forced to break its connections to former allies and getting $k_x=1$. Second, to counteract the depletion of links that constitute the random network, all individuals are allowed to form a new link with a randomly chosen player with which they are not yet connected after every $\tau$ full Monte Carlo steps. The latter process can be considered as aging, and accordingly, as soon as $k_x$ reaches a threshold $k_{max}$, player $x$ dies and is replaced by a newborn having the same strategy and keeping a single randomly selected link from its predecessor, thus maintaining $k_x=1$. Within the current work $k_{max}$ was chosen large enough so as not to influence the initial random topology of players. Below presented results were obtained on networks hosting $N=10^4 - 2 \cdot 10^5$ players, for which $k_{max}=500$ has proven to be sufficiently large. It is worth mentioning that the coevolutionary process affecting the interaction network may occasionally result in detached individual players that originally formed the neighborhood of an invaded player. In such cases we relinked the detached player randomly back onto the network with a single connection. By maintaining the minimal degree equal to one, we achieve that every player has at least one neighbor at all times, which is a necessary condition for playing the game and defining the fitness of an individual. We also highlight that the proposed rule for the adaptation of interaction networks is fully strategy independent, and thus, on its own does not indirectly support cooperators by treating $C-C$, $C-D$ or $D-D$ links differently from one another. Next, we will systematically analyze the evolution of cooperation by $\tau=1$ and $\tau=500$ for all three social dilemma types. The two considered values correspond to small and large time scale separation between link deletions and additions, and as we will show, largely affect the sustenance of the cooperative strategy as well as the mechanisms responsible for it.

\section{Results}

\begin{figure}
\scalebox{0.5}[0.5]{\includegraphics{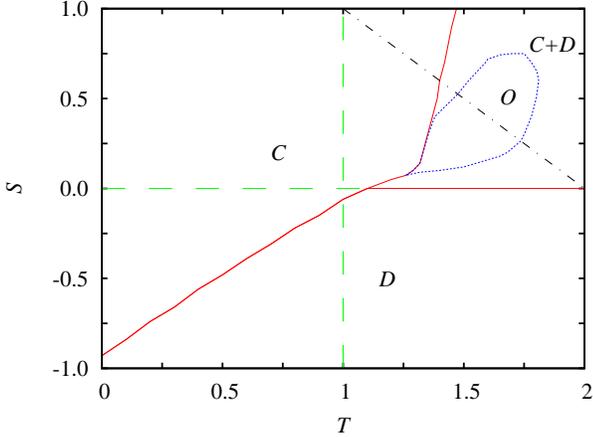}}
\caption{Full $T-S$ phase diagram obtained for $\tau=1$. Solid red lines denote the borders between full cooperator dominance ($C$ phase), full defector dominance ($D$ phase), and the mixed states ($C+D$ phase). Dotted blue line in the snowdrift quadrant (upper right) denotes the border of oscillatory solutions ($O$ phase). We further examine this quadrant along the dash-dotted black line using the common $r$ parametrization (see text for details). Dashed green lines depict borders between the social dilemma types. Note that the dominance of cooperators in the upper left quadrant is trivial since $T \leq 1$ and $S \geq 0$ simultaneously.}
\label{fig2}
\end{figure}

We start by presenting results obtained for $\tau=1$ in Fig.~\ref{fig2}, where the complete $T-S$ phase diagram is presented. In the bottom right quadrant, where the $T>R>P>S$ ranking constitutes the prisoner's dilemma game, the cooperators can be sustained up to $T=1.1$ but only if $S=0$, whereas for all $S< -0.037$ the defectors dominate completely. Moving to the stag-hunt game in the bottom left quadrant ($R>T>P>S$) of Fig.~\ref{fig2}, the sustenance of cooperators improves, yet in both cases cooperation levels do not notably surpass those that can be obtained on random networks in the absence of coevolutionary rules (see \textit{e.g.} Fig. 2 in \cite{santospnas06}). Indeed, the most interesting features can be observed in the upper right quadrant, where besides complete cooperator dominance for low enough $T$, there exist a broad region of mixed $C+D$ states including oscillatory solutions (marked with $O$). Evidently, the dynamical adaptations of the random network can sustain cooperators across the whole snowdrift quadrant, which is a notable improvement if compared to the case of static interaction networks.

\begin{figure}
\scalebox{0.5}[0.5]{\includegraphics{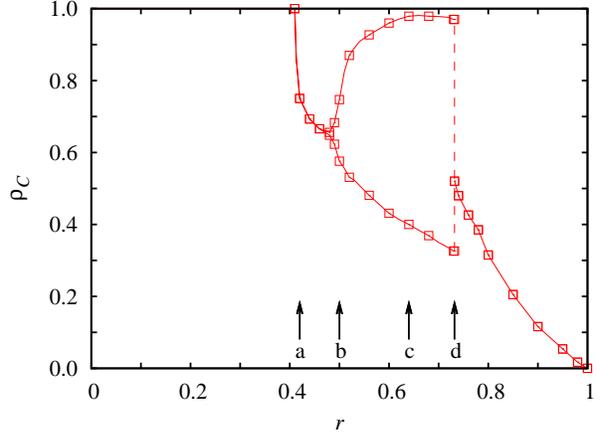}}
\caption{Density of cooperators $\rho_C$ in dependence on $r$ (the dash-dotted diagonal splitting the upper right quadrant in Fig.~\ref{fig2}) for $\tau=1$. Arrows mark sampling values of $r$ for which temporal curses of $\rho_C$ are presented in Fig.~\ref{fig4}. Note that by oscillatory solutions, starting at $r=0.50$ (arrow b) via a second order continuous phase transition and ending at $r=0.732$ (arrow d) via a first order discontinuous phase transition, both the minima and maxima of $\rho_C$ are depicted. The vertical dashed connecting lines at $r=0.732$ indicate that these states are unstable and transient either to the oscillatory or the steady state solution [see Fig.~\ref{fig4}(d) and the pertaining main text].}
\label{fig3}
\end{figure}

\begin{figure}
\scalebox{0.5}[0.5]{\includegraphics{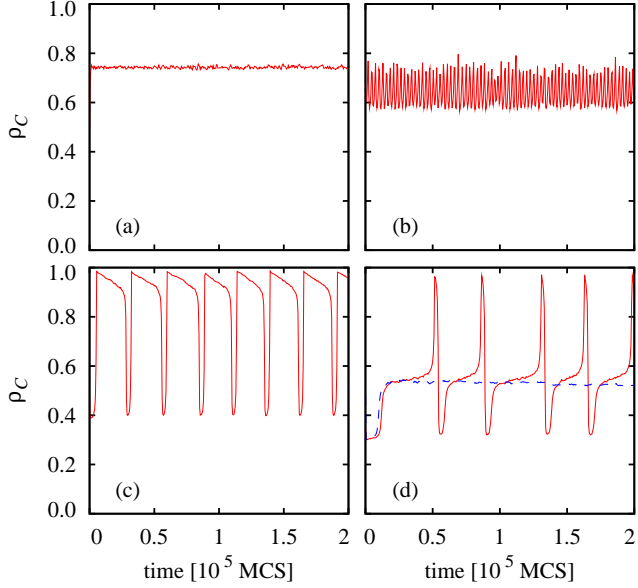}}
\caption{Temporal evolutions of $\rho_C$ for $r$ corresponding to the arrows in Fig.~\ref{fig3}; (a) $r=0.42$, (b) $r=0.50$, (c) $r=0.64$ and (d) $r=0.732$. In panel (d) the coexistence of a steady state (dashed blue line) and the oscillatory solution (solid red line) at the first order discontinuous phase transition is demonstrated (see text for details).}
\label{fig4}
\end{figure}

Aiming to identify the mechanism behind the promotion of cooperation, we analyze the evolutionary processes in the upper right quadrant of Fig.~\ref{fig2} more accurately. Without loosing generality we can introduce a single parameter $r \in (0,1)$ that determines the free payoff elements as $T=1+r$ and $S=1-r$ \cite{wangpre06}. Accordingly, the values of $r$, constituting the dash-dotted diagonal in Fig.~\ref{fig2}, characterize the cost-to-benefit ratio of the snowdrift game \cite{santosprl05}. In Fig.~\ref{fig3} we present the density of cooperators $\rho_C$ along $r$. Since the $r$ diagonal cuts through the oscillatory solutions (dotted blue region in Fig.~\ref{fig2}), we plot in Fig.~\ref{fig3} minima and maxima of $\rho_C$ for these particular values of $r$. Following complete cooperator dominance up to $r \cong 0.41$ and a rather sharp descent of $\rho_C$ (region around arrow a), the oscillatory solutions start via a second order continuous phase transition at $r \cong 0.50$. The amplitudes of these oscillations are therefore initially modest (arrow b), but increase fast beyond the transition point (arrow c). As $r$ increases further the oscillations terminate abruptly via a first order discontinuous phase transition at $r=0.732$ (arrow d), and again settle onto a stationary state. Temporal behaviors of $\rho_C$ at these characteristic values of $r$ are depicted in Fig.~\ref{fig4}.

\begin{figure}
\scalebox{0.5}[0.5]{\includegraphics{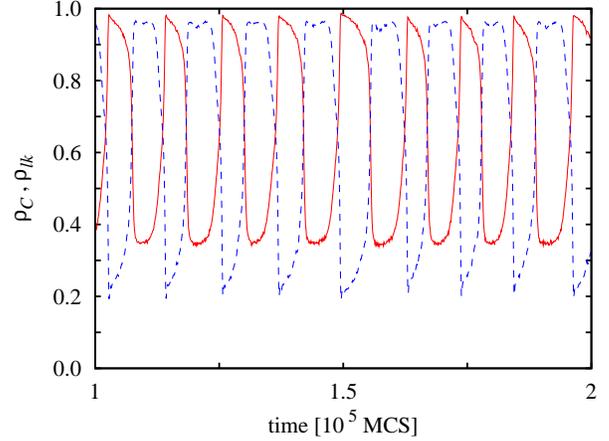}}
\caption{Temporal evolutions of $\rho_C$ (solid red line) and the density of players belonging to the low-degree class $\rho_{lk}$ (dashed blue line) obtained for $r=0.7$ (equivalently $T=1.7$ and $S=0.3)$ and $\tau=1$. Note the anti-phase synchronization between $\rho_C$ and the fraction of players having small degree.}
\label{fig5}
\end{figure}

From the dynamical point of view particularly notable is Fig.~\ref{fig4}(d), where the coexistence of steady state and oscillatory solutions, characteristic for some first order discontinuous phase transitions, is depicted. We note here that the exact sequence of bifurcations is difficult to determine in the absence of a low-dimensional model or the use of coarse-graining techniques (see \cite{grossepl08} for a related study). The basin of attraction of the steady state solution is caught if the initial value of $\rho_C$ is close enough to the stationary value (initially $\rho_C=0.6$; dashed blue), but otherwise the oscillatory solution is chosen (initially $\rho_C=0.2$; solid red). From the viewpoint of strategy evolution, results presented in Figs.~\ref{fig3} and \ref{fig4} outline a Red Queen mechanism for the sustenance of cooperation in the snowdrift game, similarly as has been reported previously if a third strategy, such as loners, was introduced to evolutionary games (see \textit{e.g.} \cite{hauertsci02, semmannnat03}). Since oscillations are possible only if the system has at least two independent variables, presently the obvious candidate for the second, first being $\rho_C$, is the network structure. Indeed, by defining $\rho_{lk}$ as the density of players whose degree belongs to the lower class (e.g. $k < (k_{max}/3)$) and measuring it in dependence on time, it can be verified that it changes with the same oscillation frequency as $\rho_C$ but a ${\rm \pi/2}$ phase shift, as shown in Fig.~\ref{fig5}. We thus argue that the tides of cooperation are evoked by the emergence of robust cooperative clusters that can form around players having high degree, similarly as reported previously for scale-free networks \cite{santosprl05}. However, since according to the coevolutionary rule the links of these players are dissolved as soon as they reach $k_{max}$, the cooperative clusters disintegrate, in turn enabling the defectors to win and thus resulting in the downfall of $\rho_C$. Simultaneously, the density of players having small(large) degree rises(falls), as evidenced by the dashed blue line in Fig.~\ref{fig5}. The process starts anew when frequent additions of new links due to a low $\tau$ value temporarily restore the difference between influential players having degree close to $k_{max}$ and the followers who have comparatively low degree, thus inducing a new oscillation cycle.

\begin{figure}
\scalebox{0.5}[0.5]{\includegraphics{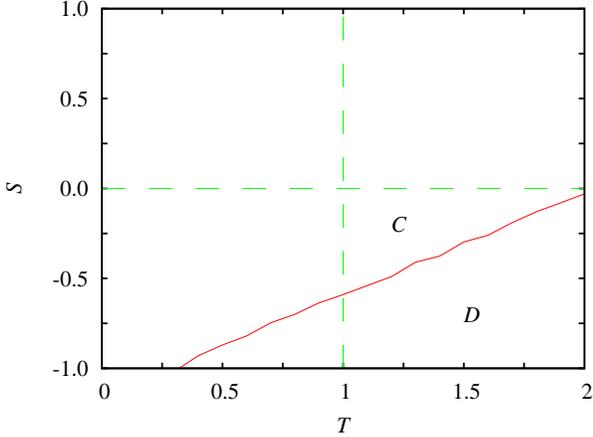}}
\caption{Full $T-S$ phase diagram obtained for $\tau=500$. Solid red line denotes the sharp border between full cooperator dominance ($C$ phase) and full defector dominance ($D$ phase). Dashed green lines depict borders between the social dilemma types.}
\label{fig6}
\end{figure}

Remarkably, when the time scale separation between link deletions and additions increases, the strength of cooperation promotion also increases substantially, as evidenced in Fig.~\ref{fig6} obtained for $\tau=500$. Note that the complete dominance of cooperators ($\rho_C=1$) now spans over an extensive $T-S$ region, encompassing the entire traditional snowdrift ($0 \leq r \leq 1$; see \textit{e.g.} \cite{santosprl05}) and weak prisoner's dilemma ($1 \leq T \leq 2$, R=1, P=S=0; see \textit{e.g.} \cite{nowaknat92}) parametrization. Moreover, the stag hunt dilemma is resolved very satisfactory as well, with defectors dominating only by extremely harsh conditions that are characterized with high $T$ and extensively negative $S< -0.6$. If compared to results presented in Fig.~\ref{fig2} ($\tau=1$), the improvement in terms of cooperation promotion in Fig.~\ref{fig6} is obvious, and it can also be observed that the mixed ($C+D$) and oscillatory ($O$) phases vanish. We thus conclude that the mechanism responsible for the promotion of cooperation changes as well, which may be attributed to two crucial difference compared to the $\tau=1$ case. First, due to the slower addition of new links the cooperative domains can grow larger and stronger around players having high degree since they prevail over longer periods of time, \textit{i.e.} the $k_{max}$ threshold is not reached so fast. And second, the slow additions of new links facilitate the emergence of highly influential players, with comparatively very high degrees if compared to the majority, to which the followers cannot catch up easily. These two facts result in a transition from the Red Queen mechanism of cooperation promotion by the snowdrift game and the predominantly heterogeneity based promotion of cooperation in the prisoner's dilemma and the stag hunt game at $\tau=1$, to a powerful group selection mechanism that emerges by all three social dilemmas for higher $\tau$, as we will demonstrate next.

\begin{figure}
\scalebox{0.5}[0.5]{\includegraphics{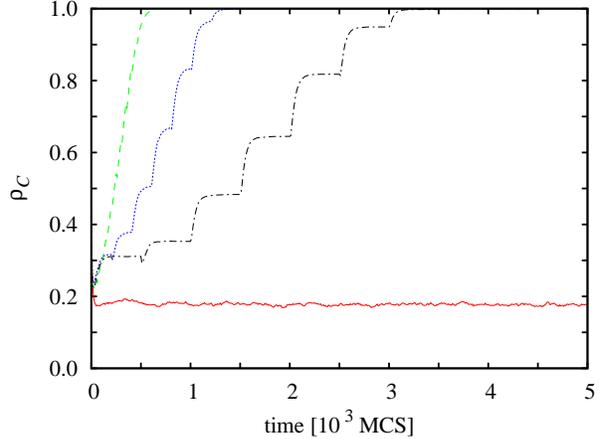}}
\caption{Temporal evolution of $\rho_C$ at $r=0.85$ (equivalently $T=1.85$ and $S=0.15)$ for $\tau=1$ (solid red line), $\tau=50$ (dashed green line), $\tau=200$ (dotted blue line) and $\tau=500$ (dash-dotted black line). Steps in the temporal evolution (time intervals during which $\rho_C$ is constant) that can be observed by larger values of $\tau$ indicate the emergence of group selection in the snowdrift game. Qualitatively identical results can also be obtained for the other two considered social dilemma types.}
\label{fig7}
\end{figure}

Figure~\ref{fig7} shows temporal courses of $\rho_C$ from the snowdrift quadrant at $r=0.85$ for increasing values of $\tau$. Most importantly, we point out the emergence of time intervals during which $\rho_C$ is constant. This cascade-like feature becomes increasingly pronounced as $\tau$ increases. Namely, at $\tau=1$ (solid red line) and $\tau=50$ (dashed green line) it is practically absent, whereas at $\tau=500$ (dash-dotted black line) the cumulative duration of dormancy of $\rho_C$ surpasses that of active phases. We argue that the reason for the emergence of these time intervals of inactivity lies in the introduced coevolutionary process, which if $\tau$ is sufficiently large, meaning that new links are added slowly, leads to the emergence of homogeneous and virtually isolated groups of players. These groups remain inactive for as long as it takes for the newly added links to reconnect them with one another, of which duration is roughly equivalent to $\tau$ (see Fig.~\ref{fig7}). It is important to note that during the inactive phase there are practically no strategy transfers taking place, and thus the main source of link deletions is disabled. Consequently, the addition of new links can gradually reconnect the detached groups, which then again triggers an avalanche of strategy adoptions, which in turn starts the whole process again, until an absorbing state is reached. Thus, the temporal plots in Fig.~\ref{fig7} arguably evidence the spontaneous emergence of group selection due to the introduction of the proposed coevolutionary rule by high values of $\tau$. Moreover, since the slow additions of new links facilitate the emergence of highly influential players, the cooperators flourish within the isolated groups whereas defectors weaken. Note that influential defectors are exposed to a negative feedback effect that sets in as soon as the neighbors adopt the defecting strategy. Then there is nobody left to exploit, and such clusters become vulnerable to cooperators. Thus, as soon as the two types of groups reestablish a sufficiently strong interconnectedness, cooperators can successfully invade the defectors, thereby gradually increasing the cooperative domains. The latter processes manifest as rather steep jumps in the temporal traces of $\rho_C$ by large enough $\tau$, which are then again followed by dormancy since the many strategy adoptions anew lead to isolation of homogeneous groups of players and restart the outlined group selection mechanism.

\section{Summary}
In sum, we show that evolving random networks constitute an optimal environment for the evolution of cooperation provided the time scale separation between the deletions and additions of links is large enough. A powerful group selection mechanism then emerges spontaneously, which is able to warrant full cooperator dominance across an extensive $T-S$ region covering all major types of social dilemmas. However, if the additions of new links are frequent, the formation of isolated homogeneous groups is hindered and the sustenance of cooperators relies either on the heterogeneity of random interaction networks, or as demonstrated for the snowdrift game, on a Red Queen mechanism which emerges due to the interplay between the oscillatory changes of the network structure and the density of cooperators. The complexity of this process is underlined by the first order discontinuous phase transition responsible for the termination of the oscillatory phase that postulates coexistence of steady state and oscillatory solutions which are chosen in dependence on the initial conditions, thus indicating the option of bistability in coevolutionary games. Notably, the existence of multiple equilibria has recently been reported also for $N$-person stag hunt dilemmas \cite{pachecoprsocb09}, whereas oscillatory solutions in the context of evolutionary games have been presented in \cite{dupa09}. Simple strategy independent coevolutionary rules, entailing both additions of new as well as deletions of existing links, can thus favor cooperative behavior beyond the levels of static complex networks, and seem to offer a rich plethora of mechanisms to tackle this formidable challenge. Thereby processes such as making new friends, punishment or aging, which may all account for the adaptations of interaction networks, are an integral part of everyday life, and it therefore seems natural to incorporate them into the framework of evolutionary game theory in order to aid the evolution of cooperation.

\acknowledgments
The authors acknowledge support from the Hungarian National Research Fund (grant K-73449), the Slovenian Research Agency (grants Z1-9629 and Z1-2032), and the Slovene-Hungarian bilateral incentive (grant BI-HU/09-10-001). Productive discussions with Gy{\"o}rgy Szab{\'o} are also gratefully acknowledged.

\end{document}